\documentclass[aps,prl,twocolumn,floatfix,superscriptaddress,showpacs]{revtex4-1}
\usepackage{graphicx}
\usepackage{amssymb}
\usepackage{amsmath}

\newcommand\pdag{{\vphantom \dagger}}
\newcommand\rs[1]{{\scriptscriptstyle\rm #1}}

\begin{document}

\title{Proposal for an optical laser producing light at half the Josephson
frequency}

\author{Frans Godschalk}
\affiliation{Kavli Institute of Nanoscience, Delft University of Technology,
P.O.  Box 5046, 2600 GA Delft, The Netherlands}

\author{Fabian Hassler}
\affiliation{Instituut-Lorentz, Universiteit Leiden, P.O. Box 9506, 2300 RA
Leiden, The Netherlands}
\affiliation{Kavli Institute of Nanoscience, Delft University of Technology,
P.O.  Box 5046, 2600 GA Delft, The Netherlands}

\author{Yuli V. Nazarov}
\affiliation{Kavli Institute of Nanoscience, Delft University of Technology,
P.O.  Box 5046, 2600 GA Delft, The Netherlands}

\date{May 2011}

\begin{abstract}
We describe a superconducting device capable of producing laser light in
the visible range at half of the Josephson generation frequency with the
optical phase of the light locked to the superconducting phase difference.
It consists of two single-level quantum dots embedded in a \textit{p-n}
semiconducting heterostructure and surrounded by a cavity supporting a
resonant optical mode.  We study decoherence and spontaneous switching in
the device.
\end{abstract}


\maketitle

Lasers and superconductors are both systems with macroscopic quantum
coherence. In lasers, photons form a coherent state induced by stimulated
emission of a driven system into a cavity mode. The resulting visible
coherent light is characterized by an optical phase \cite{scully}. In
superconductors, the ground state arising from spontaneous symmetry
breaking is also characterized by a phase \cite{tinkham}.

Traditionally, lasers and superconductors are studied separately.  Recently
\cite{recher:10}, it has been realized that the superconducting (SC)
phase difference and the optical phase may interact in a single device that
combines two superconductors and a semiconducting \textit{p-n} junction. The
latter is a common system for light generation as the electron-hole
recombination produces photons of visible frequency \cite{weisbuch}.
Combining semi- and superconductors within a nanostructure has been a
difficult technological problem that attracted attention for a long time
\cite{Nitta}. It has been solved using semiconductor nanowires \cite{Leo1}
or quantum wells \cite{Asano}, opening up the possibility to make combined
devices.

\begin{figure}[b]
  \centering
  \includegraphics{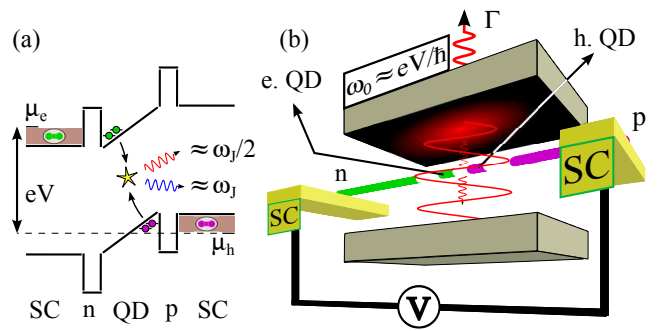}
  \caption{%
  (Color online) (a) The Josephson LED: electron (above) and hole (below)
  QD levels are close to the chemical potentials $\mu_{e,h}$ of the SC
  leads which differ by an energy $eV$. Charge transfer is only possible
  either through electron-hole recombination with the emission of a ``red''
  photon at $\tfrac{1}{2}\omega_J$ or through a Cooper pair transfer with
  the emission of a ``blue'' photon at $\omega_J$.
  (b) The HJL is a Josephson LED embedded in an optical cavity with a
  resonance frequency $\omega_0\approx eV/\hbar$, i.e., close to the
  ``red'' emission frequency. The separately-colored regions in between 
  the depleted areas represent the two QDs.
  }\label{fig:setup}
\end{figure}

The device in question has been termed a Josephson LED
[Fig.~\ref{fig:setup}(a)]. It employs a double quantum dot (QD)
in a \textit{p-n} semiconductor nanowire connected to SC leads
\cite{recher:10}. The device, biased with a voltage $V$, exhibits two
types of photon emission: ``blue'' photons at the Josephson frequency
$\omega_J=2eV/\hbar$ due to the recombination of a Cooper pair from each
side of the junction, and ``red'' photons at about $\omega_J/2$ due to
electron-hole recombination.  It has been shown that the optical phase
of the Josephson generated ``blue'' photons is locked with the SC phase
difference. The resulting ``blue'' light could in principle be enhanced 
by traditional optical methods but its small intensity makes this a 
challenging task.

In this Letter, we explore an alternative idea where the far more 
intense ``red'' emission is enhanced in a resonant cavity mode. 
We find lasing at half the Josephson frequency and, thus, dub
the device `Half-Josephson Laser' (HJL). In a common laser, lasing results
from spontaneous symmetry breaking where all values of the optical phase
are equivalent. Drift between these values leads to a finite decoherence
time. In contrast, the optical phase of the HJL is locked to the SC phase
difference with only two allowed values of the optical phase corresponding
to two opposite radiation amplitudes. This removes drift as a source of
decoherence and opens up the possibility to manipulate the optical phase by
changing the SC phase difference.  Instead, decoherence of the radiation
in the HJL results from switching between different QD states accompanied
by the emission of a photon. We have explored these processes and find
that by order of magnitude the resulting decoherence time is the same as
the theoretical limit for a common laser $\tau_{{\rm dec}} = n/\Gamma$,
with $\Gamma$ the damping rate and $n$ the number of photons accumulated
in the resonant mode. A rather low $\Gamma$ is required to achieve
lasing for a single Josephson LED, this condition being relaxed with a
large number of LEDs in a single cavity \cite{supplementary}.

\textit{Setup and model} The HJL is a Josephson LED embedded in a
single mode optical cavity with resonance frequency $\omega_0 \approx
\tfrac{1}{2}\omega_J$, Fig.~\ref{fig:setup}(b). The light emission from
the cavity is described by a damping rate $\Gamma$. The electronic part
consists of a biased \textit{p-n} junction where each side of the junction
accommodates a QD connected to a SC lead. The barriers separating the
QDs from the leads are arranged such as to allow charge transfer only
through electron-hole recombination. Such QD junctions can be realized
with semiconducting nanowires \cite{minot:07}.

The minimal model for the QDs involves a single orbital for each QD. An
orbital can house up to two particles (including spin) yielding 16 possible
states. The QD Hamiltonian then reads \cite{hassler:10}
\begin{equation}\label{eq:ham_dot}
  \hat H_\rs{QD} = \sum_{i= e,h} [ E_i \hat n_i + U_i \hat n_i (\hat n_i - 1) ]
   + U_{eh} \hat n_e \hat n_h,
\end{equation}
where $\hat n_e = \sum_\sigma \hat c^\dag_\sigma\hat c^\pdag_\sigma$
($\hat n_h = \sum_\sigma \hat h^\dag_\sigma\hat h^\pdag_\sigma$) is the
electron (hole) number operator, and $\hat c_\sigma$ ($ \hat h_\sigma$)
is the annihilation operator for an electron (hole) with spin $\sigma$.
The energies $E_{e,h}$ are measured with respect to chemical potentials
$\mu_{e,h}$ of the corresponding leads that differ by an energy $eV= \mu_e
-\mu_h$; here, $U_{e,h}>0$ is the on-site charging energy and $U_{eh}<0$
the Coulomb attraction between electrons and holes.  For concreteness, we
assume that the hole level houses a heavy hole with $J_z = \pm \tfrac{3}{2}
\hbar$, where $z$ is the nanowire axis \cite{weisbuch,niquet}.  Such levels
are commonly used in optical experiments with QDs \cite{modern_heavy_hole}.
Our qualitative results do not depend on this particular choice.

Due to the proximity of the SC leads, Cooper pairs can coherently tunnel
between the SC leads and the QDs introducing mixing between unoccupied and
doubly occupied QD states. These processes can be compactly described by an
additional term $\hat H_\rs{SC} = \tilde\Delta_e^*\hat c_\uparrow\hat
c_\downarrow + \tilde\Delta_h\hat h_\uparrow\hat
h_\downarrow + \text{H.c.}$ in the Hamiltonian; here, the induced
pair potentials $\tilde\Delta_{e,h}$ have reduced magnitudes in comparison
with the gaps $\Delta_{e,h}$ of the SC leads but they retain the same
phases $\phi_{e,h}$. Owing to gauge invariance, the physical quantities
depend only on SC phase difference $\phi \equiv \phi_e -\phi_h$
The Hamiltonian is valid under the conditions $|\tilde{\Delta}_{e,h}|,
E_{e,h}, U_{e,h}, U_{eh} \lesssim |\Delta|$.  We note further that this
Hamiltonian along with the electron-hole recombination conserves parity
(\emph{even} or \emph{odd}) of the total number of particles on the QDs.
Even-odd transitions require creation of quasiparticle excitations in the
SC leads and occur with a relatively slow rate estimated below.

Interaction between the resonant mode and QDs is described by $\hat
H_\text{int} = -{\bf E}\cdot{\bf \hat d}$; ${\bf E}$ being electric field
of the mode at QD position and ${\bf \hat d}$ the dipole moment of the
optical transition between the conduction and the valence band. We assume
a linear polarized mode, choose the $x$-axis in the direction of the
polarization, and notice that for heavy holes $\hat d_x \propto (\hat{x}
e^{-ieVt/\hbar}+{\rm H.c.})$ with $\hat{x} \equiv (\hat h_\downarrow\hat
c_\uparrow +\hat h_\uparrow\hat c_\downarrow)$. The time-dependence of
the dipole moment is due to the applied voltage. It is convenient to
implement a rotating-wave approximation transferring the time-dependent
factor to the photon creation (annihilation) operator $\hat{b}^\dagger$
($\hat{b}$). Thereby, the photon-dependent part of the Hamiltonian reads
\begin{equation}
\hat{H}_\text{ph} = \hbar \omega \hat{b}^\dagger \hat{b} + 
 G (\hat{b}^\dagger \hat{x} + \hat{b}\hat{x}^\dagger) 
\label{eq:photon}
\end{equation}
with $\omega$ being the frequency \emph{detuning}, $\omega =
\omega_0-eV/\hbar$, $|\omega| \ll \tfrac{1}{2} \omega_J$. We see that
$\hat x$ plays the role of a driving force that excites the oscillations
in the mode.  We note that all Hamiltonians considered conserve spin.

\textit{Semiclassics} The present model is a rather complex case of
nonequilibrium dissipative quantum mechanics. However, since we envisage a
large number of photons in the mode, we employ a semiclassical approximation
replacing $\hat b \mapsto \langle\hat b \rangle \equiv \lambda/G$. The
Hamiltonian $H_\rs{QD} + H_\rs{SC}+ H_\text{ph}$ can then be diagonalized to
obtain the spectrum $E_m(\lambda)$ and corresponding eigenstates $|m\rangle$.
The dipole strength $x_m(\lambda) \equiv \langle m|\hat{x}|m\rangle$
depends both on the radiation field $\lambda$ and the QD state $|m\rangle$. Since the
dipole strength in turn determines the evolution of the radiation field via
the evolution equation
\begin{equation}\label{eq:selfconsistent}
  \dot \lambda = -\left(i\omega + \frac{\Gamma}{2}\right)\lambda 
  - i\frac{G^2}{\hbar}x_m(\lambda),\quad x_m = \frac{\partial
  E_m}{\partial\lambda^\ast},
\end{equation}
we have to solve the system self-consistently \cite{scully}.  The
radiation field can build up as long as the energy gain rate $2\hbar
\omega_0(G^2/\hbar)\text{Im}[x_m(\lambda)/\lambda]$ due to the nanowire is
greater than the energy loss rate $\hbar \omega_0\Gamma$. With increasing
$\lambda$ the energy gain saturates till a {\it stationary state of
radiation} (SSR) with $\dot\lambda=0$ is reached at a certain radiation
amplitude $\lambda_s$.

In conventional lasers, the driving is due to a population inversion that
originates from dissipative transitions in an open system. For the HJL,
the SC drive is not dissipative by itself: only the emission of photons
from the cavity is a dissipative process. The driving originates from
coherent mixing of discrete quantum states due to the proximity of the
QD to the SC leads without any population inversion.  Thus, the driving
mechanism of the HJL is very different from that of a conventional laser.
This is why the information about the SC phase difference is preserved in
the process of driving.  The energy gain, including its sign, depends on the
difference between $\phi$ and the phase of $\lambda$.  Owing to this, the
phase of $\lambda_s$ of the SSR is locked to the SC phase difference. The
SSRs of the HJL come in pairs $\pm\lambda_s$ which is very different
from a conventional laser where only the magnitude $|\lambda_s|$ (photon
number) is fixed. We give in \cite{supplementary} analytical solutions to
Eq.~\eqref{eq:selfconsistent} for a toy two-level model.

\textit{Scales} Let us estimate the scales involved that are expected to 
yield lasing.  To simplify, we assume all characteristics of the QD spectrum 
to be of the same energy scale $E$ which is of the order $E \simeq |\tilde 
\Delta_{e,h}| \ll eV$. This assures optimal mixing of the QD states by 
superconductivity.  In a lasing state, the radiation amplitude should 
noticeably contribute to the energies of the QD states. This requires 
$|\lambda| \simeq E$. Assuming $\omega \simeq \Gamma$ we estimate from 
Eq.~\eqref{eq:selfconsistent} that this takes place
at $G \simeq \sqrt{\hbar\Gamma E}$. We will assume that $G$ is always chosen
to be of this scale. The number of photons is then estimated as $n \simeq
|\lambda|^2/G^2 \simeq E/\hbar \Gamma$.  The semiclassical approximation is
thus justified provided $\Gamma$ is sufficiently small, $\Gamma \ll E/\hbar$.

\textit{Lasing} Despite the model being minimal, it contains ten parameters
that affect the existence and characteristics of the SSRs.  To find these
characteristics, we need to evaluate the dipole moment in a given state at
given $\lambda$, and can do it separately for the states of \emph{odd} and
\emph{even} parity since they are not mixed by interactions.  Additionally,
the spin conservation splits the eight odd states into two equivalent
groups of four corresponding to total spin $\pm \tfrac{1}{2}$. For the
even states, only one of the four possible $|1_e1_h\rangle$ states,
$(\hat h^\dag_{\downarrow}\hat c^\dag_{\uparrow} +\hat h^\dag_{\uparrow}
\hat c^\dag_{\downarrow} ) |0\rangle$, couples to the field. Hence, we only
need to consider five of the eight even states as the other three are dark.

\begin{figure}
  \centering
	\includegraphics{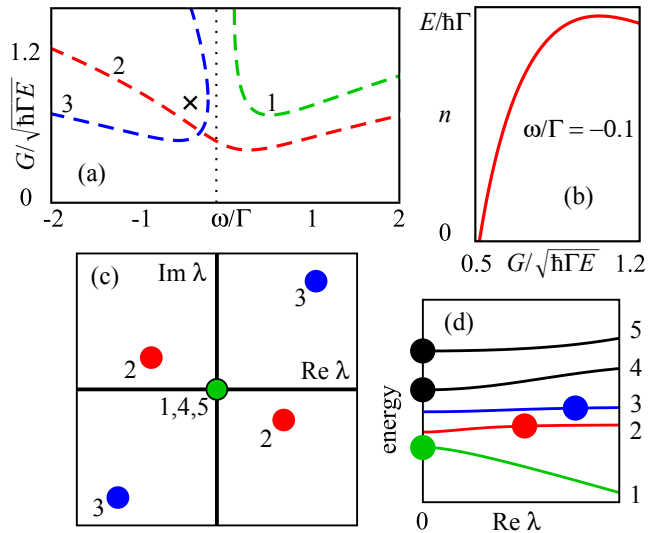}
	\caption{%
  (Color online) SSRs in the even states for QD parameters given in the
  text. The five eigenstates are labeled with numbers.
  (a) Lasing thresholds $G_c$ for three eigenstates. At the line with
  the number $m$ the energy gain at $\lambda=0$ for the state $|m\rangle$
  exactly equals the energy loss.
  (b) Number of photons $n$ for eigenstate $|2\rangle$ versus the coupling
  constant $G$ for $\omega/\Gamma=-0.1$ [dotted line in (a)] above the
  lasing threshold at $G=G_c\approx 0.5\sqrt{\hbar\Gamma E}$.
  Plots (c) and (d) illustrate the radiation amplitudes $\lambda_s^m$ (marked
  with circles) for SSRs corresponding to the different eigenstates of
  the QD.  The parameter choice is given by the cross in (a) where only the
  states $|2\rangle$ and $|3\rangle$ are lasing.  (c) Non-zero $\lambda_s^m$
  come in pairs with opposite sign. Changing the SC phase difference will
  rotate the $\lambda_s^m$ with respect to the origin of the plot.
  (d)  Eigenenergies versus $\lambda$ (at ${\rm Im}\, \lambda =0$). The 
  $\lambda_s^m$ are different for each $|m\rangle$.
  }\label{fig:stability}
\end{figure}

With this, we demonstrate lasing as proof of concept by finding SSRs 
in the even states for QD parameters within the
above estimated scales, $-E_e
= E_h = \tfrac{1}{2} U_e= \tfrac{1}{2} U_h=-U_{eh} = \Delta_h\equiv
E$ and $\Delta_e = 1.5 E$, for wide regions in the space of detuning
$\omega$ and coupling $G$, see Fig.~\ref{fig:stability}. Note that each
eigenstate $|m\rangle$ has a different dipole strength $x_m$ such that the
lasing threshold $G_c$ [Fig.~\ref{fig:stability}(a)] and the radiation
amplitudes of the SSRs $\lambda_s$ [Fig.~\ref{fig:stability}(c) and (d)]
depend on $m$.  Figure~\ref{fig:stability}(b) shows the number of photons
upon crossing the lasing threshold for the state $|2\rangle$. In agreement
with the estimations, $n$ reaches the maximum $ \simeq E/\hbar\Gamma$
at $G\simeq\sqrt{\hbar\Gamma E}$.  We stress that the optical phase of
the radiation amplitude in an SSR is not arbitrary but locked to the SC
phase difference [Fig.~\ref{fig:stability}(c)].

\textit{Switching} In the above discussion, we have assumed the QD to
stay in a certain eigenstate $|m\rangle$. In fact it does not: the finite
spectral width of the mode enables switching between the eigenstates.
As shown below, the switching events occur on a much longer timescale
$\Gamma^{-1}_{\rs{SW}}$ than that of the relaxation of $\lambda$ towards
its stationary value, $\Gamma^{-1}$.  This separation of timescales allows
to consider the switching dynamics separately from the dynamics of the
radiation amplitude.

Each switching event is accompanied by the emission of a photon with a
frequency mismatch compensating the difference of energies between initial
and final eigenstates, $\hbar\omega_k=E_f-E_i$. For switchings not altering
the parity of the eigenstate, the rates $\Gamma_\rs{SW}$ can be evaluated
using Fermi's Golden Rule that contains the effective density of photon
states $\Gamma/\hbar\omega_k^2$ (the tail of a Lorentzian-shaped emission
line of the resonant mode) and the square of the matrix element, $|\langle
m_f|\hat{H}_\text{ph} |m_i\rangle|^2$, with $|m_{i(f)}\rangle$ denoting
the initial (final) state. Thereby, the switching rate can be estimated
as $\Gamma_\rs{SW} \simeq \Gamma G^2/E^2\simeq \Gamma/n \ll \Gamma$. The
switching events are thus rare and the device stays in one of the SSRs
between the events.

Switchings altering parity are even rarer as they require the excitation
of a quasi-particle above the SC energy gap $|\Delta_{e,h}|$. The larger
detuning of the off-resonant photon $\omega_k\simeq|\Delta_{e,h}|/\hbar$
and an additional small factor $|\tilde\Delta_{e,h}/\Delta_{e,h}|$ result in
a parametrically smaller rate $\Gamma_\text{e-o}\simeq |\tilde\Delta|\Gamma
G^2/|\Delta|^3 \ll \Gamma_\rs{SW}$ \cite{recher:10}.  Such processes do
not conserve spin thus enabling switchings between dark and emitting states.

\begin{figure}[tb]
  \centering
  \includegraphics{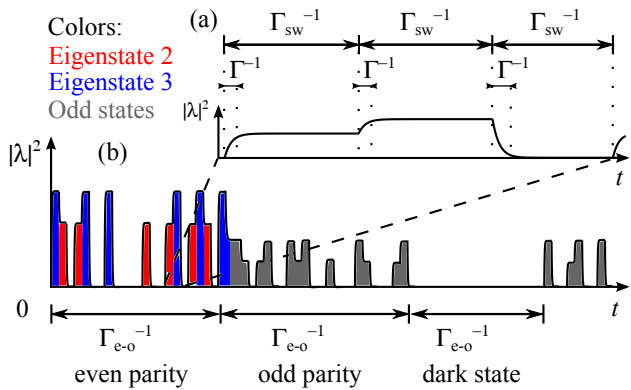}
  \caption{% 
  (Color online) Sketch of the radiation intensity $|\lambda|^2$ evolving
  in time.  (a) Switching events not altering the parity of the QDs occur
  at the timescale $\simeq\Gamma_\rs{SW}^{-1}$. After switching, the
  radiation amplitude attains its new stationary value at a timescale
  $\simeq\Gamma^{-1} \ll \Gamma_\rs{SW}^{-1} $ during which the QD
  remains in the same eigenstate. (b) Switching events altering the
  parity occur at a longer timescale $\simeq\Gamma_\text{e-o}^{-1} \gg
  \Gamma_\rs{SW}^{-1}$. These can change between the dark and lasing states.
  }\label{fig:timeline}
\end{figure}

It is important to realize that, since the SSRs for different eigenstates
have different values of $\lambda_s^m$, $\lambda$ does not jump to the
new stationary value upon a switching. Rather, the amplitude will evolve
to $\lambda_s^{m_f}$ within a timescale $\simeq \Gamma^{-1}$ , according
to Eq.~\eqref{eq:selfconsistent}.  For the same reason a switching event
always involves different eigenstates rather than different SSRs at the
same eigenstate. The latter would require large fluctuations of $\lambda$
that are suppressed exponentially.  Fig.~\ref{fig:timeline} shows a
sketch of the radiation intensity as function of time. In contrast to
common lasers, the HJL intensity fluctuations are large at timescales of
$\Gamma^{-1}_{\rs{SW}}$.

\textit{Decoherence} The intrinsic mechanism of decoherence in common
lasers is a drift of the optical phase. For the HJL, this mechanism
does not work since the amplitudes of the SSRs are locked to the SC phase
difference. This renders switching the most important source of decoherence
in the HJL. Indeed, after switching from a lasing to a non-lasing SSR the
radiation extinguishes quickly and its phase is forgotten. Even if the next
switching brings the system to a lasing eigenstate, the radiation will evolve
from the initial $\lambda=0$ to any of the two possible $\pm \lambda_s^m$,
with equal probability. Since decoherence is due to switching, the relevant
timescale is given by $\tau_\text{dec} \simeq \Gamma_\rs{SW}^{-1}\simeq
n/\Gamma$. Despite the very different decoherence mechanism, this estimation
is the same as for the common laser \cite{scully:67}.

\textit{Average power and current} The intensity fluctuations due to
switching self-average at timescale exceeding $\Gamma_{\rs{e-o}}^{-1}$. The
averaged characteristics are expressed in terms of the probabilities $P_s^m$
to be in a SSR $s$ that belongs to an eigenstate $|m\rangle$. Those are
given by the stationary solution to the master equation of the switching
dynamics, that is composed of the switching rates \cite{supplementary}. In
terms of these probabilities, the average number of photons in the cavity
is given by $\bar n = \sum_{m,s} P_s^m|\lambda_s^m|^2/G^2$. The average
emission power is proportional to the photon number, $W=\tfrac{1}{2}\hbar
\omega_J\Gamma \bar{n}$. The same holds for current in the device:
since the emission of each photon is accompanied by a charge transfer,
it is given by $I= e \Gamma \bar n = W/V$. An elaborated example of the
current/intensity dependences is provided in \cite{supplementary}.  For the
current-voltage characteristic, we find a rather complex structure beside
a peak with a magnitude of the order $e E/\hbar$, that is concentrated
in a narrow interval $\simeq \hbar\Gamma/e$ of voltages in the vicinity
$eV=\hbar\omega_0$. In this structure, two types of discontinuities are
present: (\emph{i}) kinks marking the thresholds of lasing instability at
$\lambda=0$ (second-order transitions) and (\emph{ii}) jumps signaling
the appearance of a SSR with stationary radiation amplitude $\lambda_s$
far from $\lambda=0$ (first-order transitions). We observe a relatively
high probability to remain in an SSR with a large photon number.  It is
explained from the fact that eigenstates at large $\lambda$ are close
to eigenstates of $\hat{x}$. This suppresses off-diagonal dipole-matrix
elements resulting in a suppressed rate of transitions from this state.

\textit{Feasibility} To show the feasibility of the HJL, we present here
the estimations with concrete numbers.  The SC gaps $|\Delta_{e,h}|$ are
typically $\simeq 1 \,$meV, so we can choose the QD energy scale $E\simeq
0.1 \,$meV.  To estimate the dipole strength $G \simeq ea|\mathbf{E}_0|$
, with $|\mathbf{E}_0| \simeq \sqrt{\hbar \omega_0/ \text{Vol}}$ being
the quantum fluctuation of the electric field in the mode, we assume the
cavity volume $\text{Vol}\simeq \ell^3$ with the wavelength $\ell = 2\pi
c /\omega_0 \simeq 600\,\text{nm}$, and take $a \simeq 5\,${\AA} for the
atomic distance scale. This gives the maximum $G \simeq 0.1\,$meV.

With these two values for $E$ and $G$, the minimum damping rate required
for lasing is $\Gamma \simeq G^2/\hbar E \simeq 10^{11}\,$Hz, corresponding
to quality factor $Q \simeq 10^{3}$ which is common for optical cavities.
However, in this situation the number of photons $n\simeq 1$.
This can be enhanced by increasing $Q$ and simultaneous decreasing $G$
so it remains $\simeq \sqrt{\hbar\Gamma E}$. For photonic crystal cavities
\cite{heo} quality factors $Q \simeq 10^6$ \cite{Deotare} have been measured and 
$Q \simeq 10^8$ \cite{Notomi} have been theoretically predicted. This
gives photon numbers $n \simeq 10^3$ and $n \simeq 10^5$, respectively.
The estimations of the emitted power and current at the peak do not
depend on the choice of $\Gamma$ and are given by $W \simeq 10\,$nW,
$I \simeq 10\,$nA. The requirements on $\Gamma$ can be eased and $n$
enhanced by putting many Josephson LEDs in
the same cavity. Furthermore, this also increases the emission power $W$
and the current $I$ \cite{supplementary}.

In conclusion, we have demonstrated the feasibility of generating coherent
visible light at half the Josephson frequency in a SC nanodevice. The
workings of the device resemble the spontaneous parametric down-conversion
in nonlinear optics \cite{spdc} with the superconductors playing the role
of coherent optical input. The novel driving mechanism results in locking
between optical phase and SC phase difference. The decoherence of the
emitted light originates from the switchings between different quantum
states of the device.

\acknowledgments

We acknowledge fruitful discussions with N. Akopian, L.P.  Kouwenhoven,
M. Reimer, and T. van der Sar and financial support from the Dutch Science
Foundation NWO/FOM.

\clearpage
\appendix
\widetext

\section{Supplementary Material to `Proposal for an optical laser producing light at half the Josephson frequency'}

This material consists of three parts. In the first part we show
explicitly how to find stationary values $\lambda_s^m$ of the SSRs from
Eq.~\eqref{eq:selfconsistent} using a simple toy two-level model of the
QD. In the second part, we provide a detailed example of the averaged
quantities, lasing intensity and current, in dependence on voltage as
promised in the main text. Finally, we briefly consider a device made by
placing many Josephson LEDs into the same optical cavity.

\section{Toy two-state model}

A presentational problem with the setup described in the main text, is
a relatively large number of QD states even when using all conservation
laws ($4$ states for the odd parity or $5$ states for the even parity). To
circumvent this problem and still discuss the essential properties of the
self-consistency equation and the novel driving mechanism, let us consider
a toy two-state model. The great advantage of this approach is that all
calculations can be performed explicitly.

\textit{Two-level system} For the toy model, we take one of the states to
be $|1_e 1_h\rangle$, which is unaffected by the superconductivity in the
leads. The other state is chosen as a superposition of two states with
an even number of particles in each dot, $\cos\theta |0_e 0_h\rangle +
\sin\theta\, e^{i\phi} |2_e 2_h\rangle$. These number states are mixed
by the SC leads with an angle $\theta$. The phase $\phi$ is the SC phase
difference. The Hamiltonian $H= H_\text{QD} + \lambda^* \hat x + \lambda
\hat x^\dag$ in this subspace is of the form
\begin{equation}\label{eq:ham}
  \hat H= 
  \begin{pmatrix}
     E & \cos\theta \, \lambda + \sin\theta \, e^{i\phi} \lambda^* \\
     \cos\theta \,\lambda^* + \sin\theta \, e^{-i\phi} \lambda & -E
   \end{pmatrix}.
\end{equation}
Here, $2E$ is the energy difference between the two states in the absence
of radiation and $\lambda = G \langle \hat b\rangle$ is due to the dipole
coupling to the resonant mode. Diagonalizing this Hamiltonian yields the
eigenenergies
\begin{equation}\label{eq:energy}
  E_\pm(\lambda) = \pm \sqrt{E^2 +|\lambda|^2 + \sin(2\theta) 
  {\rm Re}[\lambda^2 e^{-i \phi}]},
\end{equation}
where `$\pm$' labels the two eigenstates of Eq.~\eqref{eq:ham}. Note that for
$\lambda=0$ we have $|+\rangle =|1_e 1_h\rangle$ and $|-\rangle=\cos\theta
|0_e 0_h\rangle + \sin\theta\, e^{i\phi} |2_e 2_h\rangle$. The average
values of the dipole operator in the eigenstates are calculated as
\begin{equation}\label{eq:force}
  x_\pm(\lambda) =
  \langle \pm|\hat x |\pm\rangle 
  = \frac{\partial E_\pm (\lambda)}{\partial \lambda^*} =
 \pm  \frac{\lambda + \sin(2\theta) e^{i\phi} \lambda^*}{2 \sqrt{E^2
 +|\lambda|^2 + \sin(2\theta) {\rm Re}[\lambda^2 e^{-i \phi}]}}.
\end{equation}
The dipole serves as a driving force in the self-consistency equation
which for stationary states is given by
\begin{equation}\label{eq:selfcons}
\left(i\omega + \frac{\Gamma}{2}\right)\lambda =
    - i\frac{G^2}{\hbar}x_\pm(\lambda).
\end{equation}
As was noted in the main text, a laser field can develop when the energy gain
rate $G_E=2\hbar\omega_0(G^2/\hbar)\text{Im}[x_m(\lambda)/\lambda]$ due to
the Josephson LED is larger than the energy loss rate $\hbar\omega_0\Gamma$. With
increasing $\lambda$ the energy gain will saturate because the
denominator of Eq.~\eqref{eq:force} increases, until a {\it stationary state
of radiation} (SSR), $\lambda_s$, is reached ($\dot\lambda_s=0$) and steady
state lasing is achieved. For this toy model the energy gain is given by
\begin{align}\label{eq:gain}
  G_{E\pm} = \mp \frac{G^2}{\hbar} \frac{\sin(2\theta) \sin(2\chi-\phi)}{\sqrt{E^2
 +|\lambda|^2[1 + \sin(2\theta)\cos(2\chi-\phi)]}}
\end{align}
where $\chi$ is the optical phase, $\lambda =|\lambda| e^{i\chi}$. 

\textit{Driving mechanism} In the main text we have stressed the novelty
of the HJL driving mechanism. In conventional lasers, the driving is due
to a population inversion that originates from dissipative transitions in
an open system. However, for the HJL the drive not dissipative and the
energy gain depends crucially on mixing of QD states by the induced SC
gaps without any population inversion.

In our toy-model the essence of the driving comes about through the
mixed state $\cos\theta |0_e0_h\rangle + \sin\theta\, e^{i\phi} |2_e
2_h\rangle$ which renders both $\lambda^*\langle-| \hat x|+\rangle$ and
$\lambda^*\langle+|\hat x|-\rangle$ non-zero. Therefore, irrespective of the
eigenstate of the QD, the system can always decay to the other state and emit
a photon. Without mixing ($\theta=0,\pi$) this is not possible. The energy
required for the decay is supplied by the bias. With increasing $\lambda$
the values of the amplitudes decrease as the eigenstates $|+(-)\rangle$
will become increasingly more like the eigenstates of $\hat x$. Hence the
driving saturates.

From the expression for the energy gain, we explicitly see the role of the SC
phase difference, $\phi$. First, the value of this phase determines the sign
of the energy gain, and thus whether it acts as a gain or as an absorber
of radiation.  Second, the SSRs with radiation amplitudes, $\lambda_s$,
depend on the specific combination of phases, $2\chi-\phi$. Hence, the SSR
values of the optical phase must depend on the SC phase difference, which
signifies a phase-lock between the optical phase and SC phase difference.

We note that, because of this specific combination of phases, there
are always two optical phases possible since changing $\chi\to\chi+\pi$
yields an SSR with the same radiation magnitude $|\lambda_s|$ but opposite
sign. Hence, the SSRs come in pairs $\pm\lambda_s$. This is in stark contrast
to conventional lasers where the driving is independent of the optical
phase and all phases occur with equal probability.

As a final remark concerning the drive we note that the dipole $x_{\pm}$
completely saturates in the limit of $\lambda \gg E$, to
\[
x_{\pm} \approx \pm e^{i\chi} \frac{ 1+ \sin(2\theta) e^{i\phi - 2i\chi}}
{2 \sqrt{1+ \sin(2\theta)\cos(2\chi-\phi)}}.
\]
In this limit the eigenstates of $H$ are the eigenstates of $\hat x$
so that the amplitudes $\langle - | \hat{x} | + \rangle$ and $\langle +
| \hat{x} | - \rangle$ vanish.

\textit{Stationary states of radiation} To find the radiation amplitudes of
the SSRs, $\lambda_s^{m=\pm}$, we need to solve the stationary self-consistency
equation with the dipole of Eq.~\eqref{eq:force}. Here we see by inspection
that $\lambda=0$ is always a stationary solution since the dipole is
proportional to $|\lambda|$. It may be unstable however against small
fluctuations in the mode. By performing stability analysis we distinguish
between the two cases.

To find a SSR with nonvanishing radiation amplitude, we first divide both
sides of Eq.~\eqref{eq:selfcons} (taking dipole $x_+$) with its conjugate
to derive expressions for the optical phase
\begin{equation}
e^{2i\gamma} = \frac{ 1+2i \omega/\Gamma}{ 1-2i\omega/\Gamma} 
= 
- \frac{1+\sin(2\theta) e^{i\phi - 2i\chi}}
{1+\sin(2\theta) e^{2i \chi -i\phi} }.
\end{equation}
where $\gamma = \arctan(2\omega/\Gamma)$. This reduces to
\begin{equation}\label{eq:phase}
\cos(2\chi- \phi +\gamma) 
= -\frac{\cos\gamma}{\sin(2\theta)}.
\end{equation}
with extra condition ${\rm sgn}[\sin(2\theta)\sin(2\chi-\phi)] =1$.
From this, we explicitly see that the optical phase $\chi$ may assume
two equivalent values shifted by $\pi$ and that it is locked to SC
phase $\phi$.  Nontrivial solutions for the optical phase $\chi$ exist for
$\omega <-\tfrac{1}{2}\Gamma \bigl|\cot 2\theta\bigr|$, i.e., only at
negative detuning. For the `$-$' eigenvalue, the situation is reversed
in detuning: the lasing solutions exist only at positive $\omega$ with
$\omega> \tfrac{1}{2} \Gamma \bigl|\cot 2\theta\bigr|$.

Now taking modulus square of both sides of \eqref{eq:selfcons}, we arrive
at an expression for the optical field magnitude
\begin{equation}\label{eq:modulus}
\frac{G^4}
{\hbar^2 (\omega^2+ \tfrac{1}{4}\Gamma^2)} 
= 4\frac{E^2 +|\lambda|^2[1+\sin(2\theta)\cos(2\chi-\phi)]}
{1+\sin^2(2\theta) + 2 \sin(2\theta)\cos(2\chi-\phi))}.
\end{equation}
Solving Eqs.~\eqref{eq:phase} and \eqref{eq:modulus} simultaneously yields
all SSRs, $\lambda_s^{m=\pm}$.

From Eq.~\eqref{eq:modulus} we may derive expressions for the lasing
thresholds as it implies that the lasing solutions with $|\lambda|>0$
can only exist above some critical coupling
\[
\frac{G^2_c}{\hbar} = \frac{2|E|\sqrt{\omega^2+ \tfrac{1}{4}\Gamma^2}}
{\sqrt{ 1+\sin^2(2\theta) + 2 \sin(2\theta)\cos(2\chi-\phi)}}= 
\frac{2|E|}{\cos^2 (2\theta)} 
\biggl[|\omega| - \sqrt{\sin^2(2\theta)\omega^2 -
\tfrac{1}{4}\cos^2(2\theta)\Gamma^2}\biggr].
\]
Stability analysis shows that the zero stationary solution becomes unstable
at this critical coupling. Because the lasing field continuously starts
to grow when crossing this threshold, we have the analog of a second-order
phase transition. Interestingly, at yet higher coupling
\[
\frac{G^2_{c2}}{\hbar} = 
\frac{2|E|}{\cos^2 (2\theta)} 
\biggl[ |\omega| + \sqrt{\sin^2(2\theta)\omega^2 -
\tfrac{1}{4}\cos^2(2\theta)\Gamma^2} \biggr]
\]
the zero solution becomes stable again and we encounter an analog of the
first-order transition where the SSRs with nonvanishing radiation amplitudes
still exist at the threshold. At the tricritical point
\[
G= \sqrt{\frac{2 |E| \hbar\Gamma}{|\sin (4 \theta)|}},\; 
|\omega|=\frac{\Gamma}{2} |\cot 2\theta|,
\]
the transitions of both orders coexist.

\section{Example of current/intensity dependence on voltage}

\begin{figure}  
  \centering
  \includegraphics{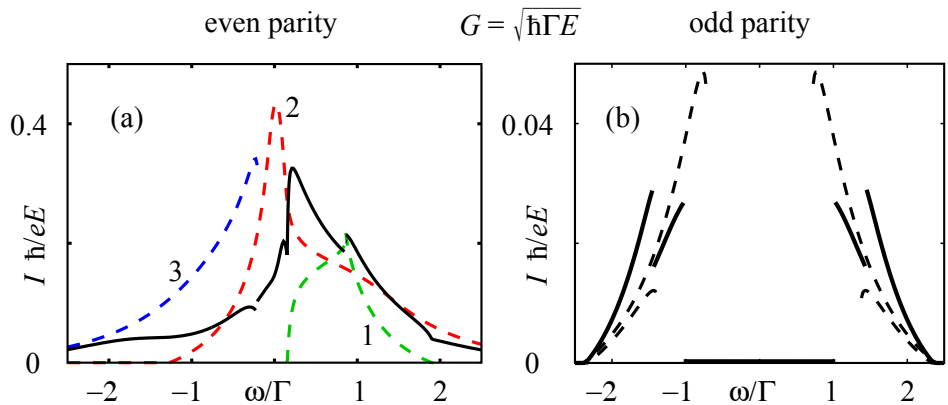}
  \caption{%
  (Color Online) Time-averaged current versus detuning for the even and
  the odd parity states. The dashed curves (reduced by factor of 3) give
  the current in individual lasing QD eigenstates, with the numbers in the
  even parity plot matching those of Fig.~\ref{fig:stability}, in the main
  text. The solid curves represent the time-averaged currents which are the
  individual currents weighted with the probabilities $P_s^m$, the stationary
  solutions to master equation~\eqref{eq:masterswitch} of the switching
  dynamics as explained in the text.  The QD parameters are as given in
  the main text, $-E_e = E_h = \tfrac{1}{2} U_e= \tfrac{1}{2} U_h=-U_{eh}
  = \Delta_h\equiv E$ and $\Delta_e = 1.5 E$, and $G = \sqrt{\hbar\Gamma E}$.
  }\label{fig:curves}
\end{figure}

In the main text the dynamics of the HJL over longer times was described
in terms of switching rates. It was argued that these switchings occur
after such long times that the corresponding dynamics is decoupled from the
dynamics of the lasing field $\lambda$. Hence, the switchings always occur
when the system is at an SSR of Eq.~\eqref{eq:selfconsistent}, $\lambda_s^m$,
where the QD is in eigenstate $|m\rangle$. We additionally argued that
(for $|\tilde\Delta|\ll |\Delta|$) switchings that change the parity of the
QD state occur at an even slower rate. Even for $|\tilde\Delta|\simeq0.1
|\Delta|$ the parity switchings $\Gamma_\text{e-o}$ are an order of
$10^3$ slower than the switchings $\Gamma_\rs{SW}$. Therefore, when
measuring over times much longer than $\Gamma_\rs{SW}^{-1}$ but shorter
than $\Gamma_\text{e-o}^{-1}$, we see an average intensity corresponding
to the SSRs of many QD eigenstates with equal parity. After a typical time
scale $\Gamma_\text{e-o}^{-1}$ the QD states switch parity and the average
intensity corresponding to this parity is seen. The purpose of this section
is to calculate the average intensity as a function of voltage for the even
and odd parities separately and discuss some important features found in
the corresponding curves.

The average intensity can be found by solving the master equation
governing the switching dynamics and finding the corresponding stationary
solution. The probability $p_s^m$ to be at $\lambda_s^m$ evolves in time
according to
\begin{equation}\label{eq:masterswitch}
	\frac{d p_s^m}{dt} =\sum_{n,r \neq m,s} 
  [ W_{m,s; n,r}\ p_r^n - W_{n,r; m,s}\ p_s^m 
	]\equiv\sum_{n,r}\mathbf{W}_{m,s; n,r}\ p_r^n
\end{equation}
where $W_{m,s; n,r}$ denotes the transition from
$\lambda_r^n\mapsto\lambda_s^m$ and we define the transition matrix
$\mathbf{W}$. The summation is over the stationary solutions
$\lambda_r$ of all eigenstates, $|n\rangle$.  To solve master
equation~\eqref{eq:masterswitch} we need to find and diagonalize $\mathbf
W$. We notice that this matrix has a left eigenvector with zero eigenvalue
because $\sum_{m,s}\mathbf{W}_{m,s; n,r}=0$. Hence there must also be a right
eigenvector with zero eigenvalue, which is the stationary distribution of
probabilities ($\dot p_s^m=0$ for all $m$) when normalized to unity. Since
we are only interested in these stationary solutions we only need to find
the right null vector of $\mathbf{W}$. We represent this stationary solution
as $P_s^m$, with a capital $P$.  Hence, we can find $P_s^m$ when all $W_{m,s;
n,r}$ are known.

To calculate the transitions $W_{m,s; n,r}$ we note that the switchings
are accompanied by the emission of an off-resonant photon so that the
rates follow from Fermi's Golden Rule. The transition rates are then
proportional to the product of $|\langle m(\lambda_r^n)|\hat{H}_\text{ph}
|n(\lambda_r^n)\rangle|^2$ and a Lorentzian shaped photon density of states,
$\rho(\Delta E) = (1/2\pi)\hbar\Gamma/[(\hbar\Gamma/2)^2+(\Delta E)^2]$, with
$\Delta E=|E_n-E_m|$ the energy difference between $|n(\lambda_r^n)\rangle$
and $|m(\lambda_r^n)\rangle$. We note that the transitions go from eigenstate
$|n(\lambda_r^n)\rangle\mapsto |m(\lambda_r^n)\rangle$ ($n\ne m$), which are the
eigenstates of the Hamiltonian $H_\rs{QD}+H_\rs{SC} + H_\text{ph}$ in the
semiclassical approximation, $\hat b \mapsto \langle\hat b \rangle \equiv
\lambda/G$. Additionally we stress that the value of $\lambda$ is unaltered
during the transition, which is indicated by the $(\lambda_r^n)$ in $\langle
m(\lambda_r^n)|$ and $|n(\lambda_r^n)\rangle$. Only after the switch will
$\lambda$ evolve to $\lambda_s^m$ according to Eq.~\eqref{eq:selfconsistent}.

To find the matrix elements of $\hat H_\text{ph}$ [Eq.~\eqref{eq:photon}]
we only need to consider the interactions $G(\hat b^\dagger \hat x + \hat
b \hat x^\dagger)$ as the term $\hbar\omega \hat b^\dagger\hat b$ will
always yield zero contribution when $n\ne m$. The two interactions imply
two possible transition amplitudes distinguishing two types of emission:
one far below and one far above the cavity resonance. Since we are in the
rotating frame, this translates to creating a negative frequency photon
with $\hat b\hat x^\dagger$, or creating a positive frequency photon with
$\hat b^\dagger\hat x$. We do not expect absorption of off-resonant photons
as they are very scarce. Taking the detuning into account, the photon
energies are $\hbar\omega_k = \hbar\omega-\Delta E$ and $\hbar\omega_k =
\hbar\omega+\Delta E$ respectively. The detuning may be neglected, however,
as it is typically taken of the same order as $\Gamma \ll \Delta E$. For
this same reason we may also neglect the $\Gamma$ in the denominator of
the density of states $\rho(\Delta E)$.

Taking the above considerations into account we find the transition rates
\begin{align}\label{eq:transition}
	W_{m,s; n,r} = \frac{\Gamma G^2}{(\hbar\omega_k)^2} | \langle m(\lambda_r^n)|\hat x|n(\lambda_r^n)\rangle - \langle m(\lambda_r^n)|\hat x^\dagger |n(\lambda_r^n)\rangle |^2.
\end{align}
With this we can find $\mathbf W$ and thus $P_s^m$. We make some brief
remarks about $\mathbf W$. As stipulated above the switchings only occur
between different QD eigenstates. Fluctuations that let the system jump
from one to another SSR in the same QD eigenstate are exponentially small
due to the large number of photons. Thus $W_{m,s;m,r}=0$ if $ s\ne
r$. Furthermore, we note that when in a lasing eigenstate and starting at
$\lambda=0$ the system may evolve to both SSRs with $\pm\lambda_s^m$ with
equal probability. Hence the transition rate from a non-lasing QD eigenstate
to the SSR with $\lambda_s^m$ of a lasing QD eigenstate is half the value
calculated in Eq.~\eqref{eq:transition}. For transitions between two lasing
eigenstates it must be determined to which of the multiple stationary
points $\lambda_s^m$ the field evolves. This is done by integrating the
time evolution equation of $\lambda$, Eq.~\eqref{eq:selfconsistent}, until
one of the $\lambda_s^m$ is sufficiently approached, using the radiation
amplitude $\lambda_r^n$ of the previous SSR as initial condition.

With the stationary probability distribution, $P_s^m$, we calculate some
measurable quantities. First, the average number of photons in the cavity
is $\bar n = \sum_{m,s} P_s^m|\lambda_s^m|^2/G^2$. Then the average number
of photons that escape the cavity is $\Gamma \bar n$, yielding the average
emission power $W=\tfrac{1}{2}\hbar\omega_J\Gamma\bar n$ and the average
current in the device $I=e\Gamma \bar n$. The latter because every photon
emission is accompanied by a single charge transfer.

Using the minimal model of the main text we calculate the current/intensity
as a function of detuning. As mentioned above we plot the dependencies
for odd and even parity states separately. The results are presented
in Fig.~\ref{fig:curves}. Since the detuning is related to the applied
voltage via $\omega = eV/\hbar-\omega_0$, the peaks in the current, of
the order of $e E/\hbar$, are in a narrow interval $\simeq \hbar\Gamma/e$
of voltages. There are two kinds of discontinuities in the curves: kinks,
that mark the thresholds of the lasing instability at $\lambda=0$ in some QD
eigenstate (second-order transitions), and jumps, that signal the appearance
of an SSR far from $\lambda=0$ (first-order transitions). The curves
also show a relatively high probability to be in lasing QD eigenstates
if available. This can be qualitatively understood by considering that
eigenstates at large $\lambda$ are close to eigenstates of $\hat{x}$. This
suppresses off-diagonal matrix elements, Eq.~\eqref{eq:transition}, resulting
in a slower rate of transitions away from a lasing state. For the odd parity
states in Fig.~\ref{fig:curves}, the time-averaged current drops to zero
while a lasing state is still available.  This is because the non-lasing
solution in this lasing eigenstate becomes stable at $|\omega|/\Gamma
\lesssim 1$ and whereas all other states are nonlasing. The device then
gets stuck in the SSRs at $\lambda=0$ even though SSRs with $\lambda_s\ne0$
are available.

\section{Many Josephson LEDs in a single cavity} 

We have discussed in the main text that the large number of photons $n
\simeq E/\hbar \Gamma$ (assuming $G\simeq \sqrt{\hbar \Gamma E}$)
in a cavity with a single Josephson LED requires high $Q$ factors. This
requirement can be easily softened, and the lasing can be achieved at
higher damping rate $\Gamma$, if we incorporate $N \gg 1$ Josephson LEDs
in the cavity. Since the LEDs are of ten nanometer scale, one can put
hundreds of them even into a single-wavelength cavity. The resulting
system certainly deserves to be explored in detail, and will be a topic
of separate research presented elsewhere. In this note we just give some
straightforward estimations.

All $N$ QD dipoles are in this setup coupled to a single mode, were we
assume equal coupling coefficients $G$.  The self-consistency equation
for the radiation amplitude changes to
\begin{equation}\label{eq:selfcons-manywires}
  \dot \lambda = -\left(i\omega + \frac{\Gamma}{2}\right)\lambda 
  - i\frac{G^2}{\hbar}\sum_i^N x_{m_i}(\lambda),\quad x_{m_i } = \frac{\partial
  E_{m_i}}{\partial\lambda^\ast}.
\end{equation}
Assuming for simplicity all dipoles to be in the same direction with same or
similar parameter values, we may simplify the sum over dipoles to $N x_m$.
This yields $\lambda \simeq G^2 N/\hbar\Gamma$ for the SSRs. The optimal
regime near the lasing threshold is achieved if $\lambda \simeq E$. Thus
this corresponds to $G \simeq \sqrt{\hbar\Gamma E/N}$. Therefore, the lasing
threshold can be achieved at much larger damping rate $\Gamma \simeq G^2
N/\hbar E$.

This also increases the number of photons in the mode.  If we
optimize the coupling to $G \simeq \sqrt{\hbar\Gamma E/N}$, the number
is estimated as $n \simeq N E/\hbar\Gamma$, $N$ times larger than the
estimate for a single-LED device.  If we fix $\Gamma$, $G$ and increase
the number of wires in the device, the photon number scales as $n \simeq
N^2 G^2/\Gamma^2$.  Note that in both cases the emitted power increases as
well. It is respectively given by $W\simeq\Gamma eV n\simeq eV NE/\hbar$,
for optimal coupling, and $W\simeq eV N^2 G^2/\Gamma$ when only increasing the
number of wires.

It is likely that a large number of LEDs in the cavity greatly reduces
the intensity fluctuations. With only one pair of QDs the lasing intensity
fluctuates down to zero as can be seen from Fig.~\ref{fig:timeline}, in the
main text. In contrast, the switching of only one of $N$ QDs will hardly
affect the intensity, so the relative intensity fluctuations are expected to
be of the order of $1/\sqrt{N}$. Likewise, the phase of the optical field
will be barely affected by the switching in one nanowire. Therefore the
decoherence times may become significantly longer than in a single LED HJL.

\end{document}